# Modeling Filamentary Conduction in Reset Phase Change Memory Devices

Md Samzid Bin Hafiz, Helena Silva, Ali Gokirmak

Electrical and Computer Engineering, University of Connecticut, Storrs, CT 06269-4157, USA

*Abstract*—. We performed a computational analysis on percolation transport and filament formation in amorphous $Ge_2Sb_2Te_5$ (a-GST) using 2D finite-element multi-physics simulations with 2 nm out-of-plane depth using an electric-field and temperature dependent electronic transport model with carrier activation energies that vary locally around 0.3 eV and as a function of temperature. We observe the snapback (threshold switching) behavior in the current-voltage (I-V) characteristics at ~50 MV/m electric field with 0.63 µA current for 300 K ambient temperature, where current collapses onto a single molten filament with ~ 2 nm diameter, aligned with the electric field, and the device switches from a high resistance state ($10^8$ Ω) to a low resistance state ($10^3$ Ω). Further increase in voltage across the device leads to widening of the molten filament. Snap-back current and electric field are strong functions of ambient temperature, ranging from ~ 0.53 µA at 200 K to ~ 16.93 µA at 800 K and ~ 85 MV/m at 150 K to 45 MV/m at 350 K, respectively. Snap-back electric-field decreases exponentially with increasing device length, converging to ~ 38 MV/m for devices longer than 200 nm.

## I. Introduction

Recent years have seen a rapid growth in data generation and processing demands due to advances in artificial intelligence, data analytics, and neuromorphic computing.[1], [2] Conventional computer architecture and memory technologies are struggling to cope with this demand[3], [4]. One of the major performance limitations is the memory access latencies between CPU, high speed low density volatile memory (DRAM), and low speed high density non-volatile memory (flash)[5], [6]. Emerging electronic memory technologies aim to fill the performance and density gap in memory hierarchy[7]–[12]. Phase change memory (PCM) has demonstrated significant prospect in this regard[2], [13] due to its faster read/write speed (10~100ns) and superior endurance (~$10^{12}$ cycles) compared to flash while maintaining similar density as flash[14] and data retention (10 years at CPU temperature)[13]–[15].

Typical PCM devices are fabricated as vertical cells using phase change materials with top and bottom contacts. PCM materials offer (i) reversible switching between their amorphous and crystalline phases, (ii) significant contrast in resistivity between their phases, (iii) fast switching between phases[2] and (iv) low thermal conductivity for power efficiency[16], [17]. Chalcogenide alloys have emerged as suitable phase change materials[18]–[22], among which $Ge_2Sb_2Te_5$ (GST) is found out to be the most promising and is the most-studied phase change material[18], [23]–[28]. GST shows high resistance contrast (~$10^4$) between its amorphous and crystalline phases[26], [29]–[31], high crystallization temperature (~400 K)[32], low melting point ($T_{melt}$ ~858 K[25]), rapid crystallization (tens of nanoseconds)[33], and excellent thermal stability[27].

PCM devices are switched using short duration voltage pulses that give rise to self-heating[34]. Amorphization (reset) is achieved by heating the active region of the cell to the melting temperature, followed by rapid quenching[35], [36]. Crystallization (set) is achieved by heating the active region above glass transition (crystallization) temperature but below $T_{melt}$ for a sufficient duration or by slower cooling from melt[37].

PCM devices experience significant local heating, extreme temperatures and thermal gradients, while the phase change material experiences nucleation and growth of crystals and amorphization. The amorphous phase change material properties are strong functions of electric field and temperature. The local variability in the disordered amorphous phase increases the complexity in operation dynamics. Hence, modeling PCM devices is more complicated compared to traditional semiconductor devices.

Here, we study electronic transport in the disordered amorphous phase of GST using an electro-thermal finite element model in 2D [38]–[42] that captures local variations in the materials parameters that depend on temperature and electric field, using COMSOL Multiphysics[43]. The local variations in the materials properties are modeled using a locally varying carrier activation energy ($E_A$) which is also a function of temperature. Electronic transport is modeled using a hyperbolic sine model that captures thermionic emission over a local barrier. The model parameters are based on the experimental values we extracted from high-speed measurements on metastable a-GST [32] (Fig. 1). The functions that define the electric-field and temperature dependence are based on electric-field and temperature dependent measurements performed on stabilized amorphous GST (a-GST) line-cells [44], [45] (Fig. 2).

## II. Electronic Transport in Amorphous GST

The resistivity versus temperature behavior of metastable a-GST (Fig. 1a) is obtained from a large number of high-speed measurements performed on GST line-cells in the 300 K to 600 K temperature range [32].

The metastable resistivity versus temperature characteristics show a simple exponential behavior which can be fitted to[46]:

$$\rho = \rho_1 e^{-\alpha T} \quad (1)$$

where $\rho_1 = 35137\ \Omega \cdot cm$ and $\alpha = 0.0202$. The resistivity can also be expressed as Arrhenius form [46]:

$$\rho = \rho_0 e^{\frac{E_A(T)}{k_B T}} \quad (2)$$

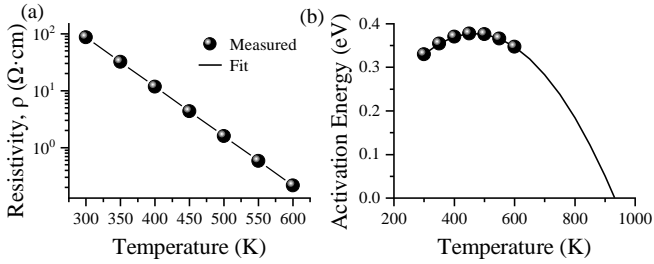

Fig. 1. (a) Measured (high speed, device level) metastable resistivity, ρ (in $\Omega \cdot cm$) of amorphous GST as a function of temperature (in K). (b) Extracted activation energy, $E_A$ (in eV) from resistivity as a function of temperature (in K).

Assuming $E_A(T_{melt})$ is equal to the thermal energy ($\frac{3}{2}k_B T$) and equating (1) and (2) at $T_{melt}$ we obtain:

$$\rho_0 e^{\frac{\frac{3}{2}k_B T_{melt}}{k_B T_{melt}}} = \rho_1 e^{-\alpha T_{melt}}$$

$$\rho_0 = \rho_1 e^{-\frac{3}{2} - \alpha T_{melt}} \quad (3)$$

From equations (1), (2) and (3), $E_A(T)$ can be written as[46]:

$$E_A(T) = k_B T \left\{ \frac{3}{2} + \alpha(T_{melt} - T) \right\} \quad (4)$$

Hence, $E_A(T)$ of a-GST has a parabolic form and reaches zero around 930 K (Fig. 1b), metal transition temperature ($T_{metal}$).

We performed additional experiments in 80 K – 350 K temperature range to characterize electric-field and temperature dependent electronic transport in a-GST. In these experiments, we used cells with width (W) × length (L) × thickness (t) ≈ 152 nm × 710 nm × 20 nm from different dies from the same wafers used in high-speed measurements[32]. After the cells were amorphized at lowest temperature (~ 80 K) using short reset pulses, high voltage DC I-V sweeps (0 V to 25 V) were performed. The very first two sweeps dramatically accelerate resistance drift and stabilize the cells (stopping resistance drift), and the consequent sweeps display stable I-V characteristics, allowing characterization of electric-field and temperature dependent electronic transport in these cells (Fig. 2).

We observe two distinct exponential responses in the low-field and high-field regimes, and the high-field stress is required to substantially accelerate resistance drift[44], [45]. The observed change in the I-V characteristic above a critical electric field (~ 20 MV/m) suggests two regimes for electronic transport which we have discussed in a recent manuscript[47]. In these experimental studies on line-cells, only a fraction of each cell is amorphized, hence the amorphous region is straddled by crystalline GST (c-GST) regions. The charge exchanges at these interfaces are expected to lead to a device level potential barrier and tunnel junctions at the interfaces. Hence, electronic transport is expected to be due to electrons being injected from the valance-band of the crystalline regions to the conduction band of the amorphous region [47]. The electronic carrier type (p or n) determines the energy exchanges in the structure (thermoelectric contributions).

In this study, we are simulating short a-GST regions (20 nm to 200 nm) with TiN top and bottom contacts. TiN is a degenerate n-type semiconductor and a-GST is a slightly p-type semiconductor. We constructed the energy-band diagram prior to junction formation (Fig. 3) for these cases using parameters obtained from the literature: for a-GST, bandgap is 0.7 eV[48]–[50], electron affinity is 4.99 eV[50], and the Fermi level is estimated to be 0.15 eV above the valance-band edge assuming a carrier concentration of 1×10$^{17}$ cm$^{-3}$ [48]; for TiN, bandgap, electron affinity, and work function are 3.2 eV, 4.25 eV, and 4.04 eV respectively[51].

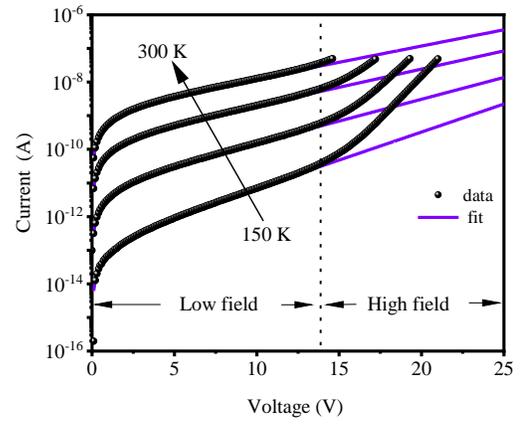

Fig. 2. I-V measurements of stable amorphous GST line cells with W × L × t ≈ 152 nm × 710 nm × 20 nm from 150K to 300K with 50K steps.

The TiN / a-GST / TiN heterostructure shows substantial potential barriers for holes and tunneling barriers for electrons at both interfaces at equilibrium condition, when the Fermi levels are aligned after the charge exchange is complete (Fig. 4a,d). The valance-band potential well that forms in a-GST is expected to contain the holes in the amorphous region even if they are activated from the hole traps[52]–[54] in the bandgap. When an bias voltage is applied, electron current is expected due to electron injection from the conduction band of TiN[55] to the conduction band of a-GST (Fig. 4b and e), through (tunneling) or above (thermionic emission) the barrier formed at the interfaces [47] (Fig. 5), and trap-assisted electronic transport through a-GST[56]. Trap-assisted electron transport can be modeled as thermionic emission over a barrier with an activation energy ($E_A$) modulated with the external field (Fig. 5) with forward and reverse transmission probabilities:

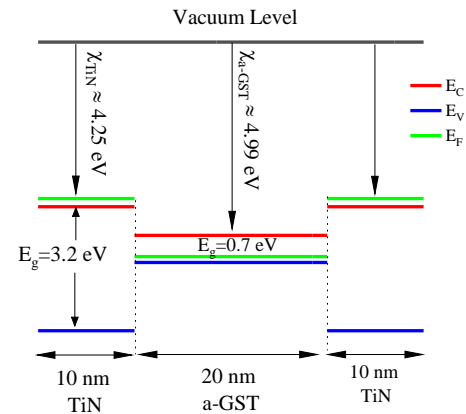

Fig. 3. Energy-band diagrams of a-GST and TiN before junction formation [48]–[51].

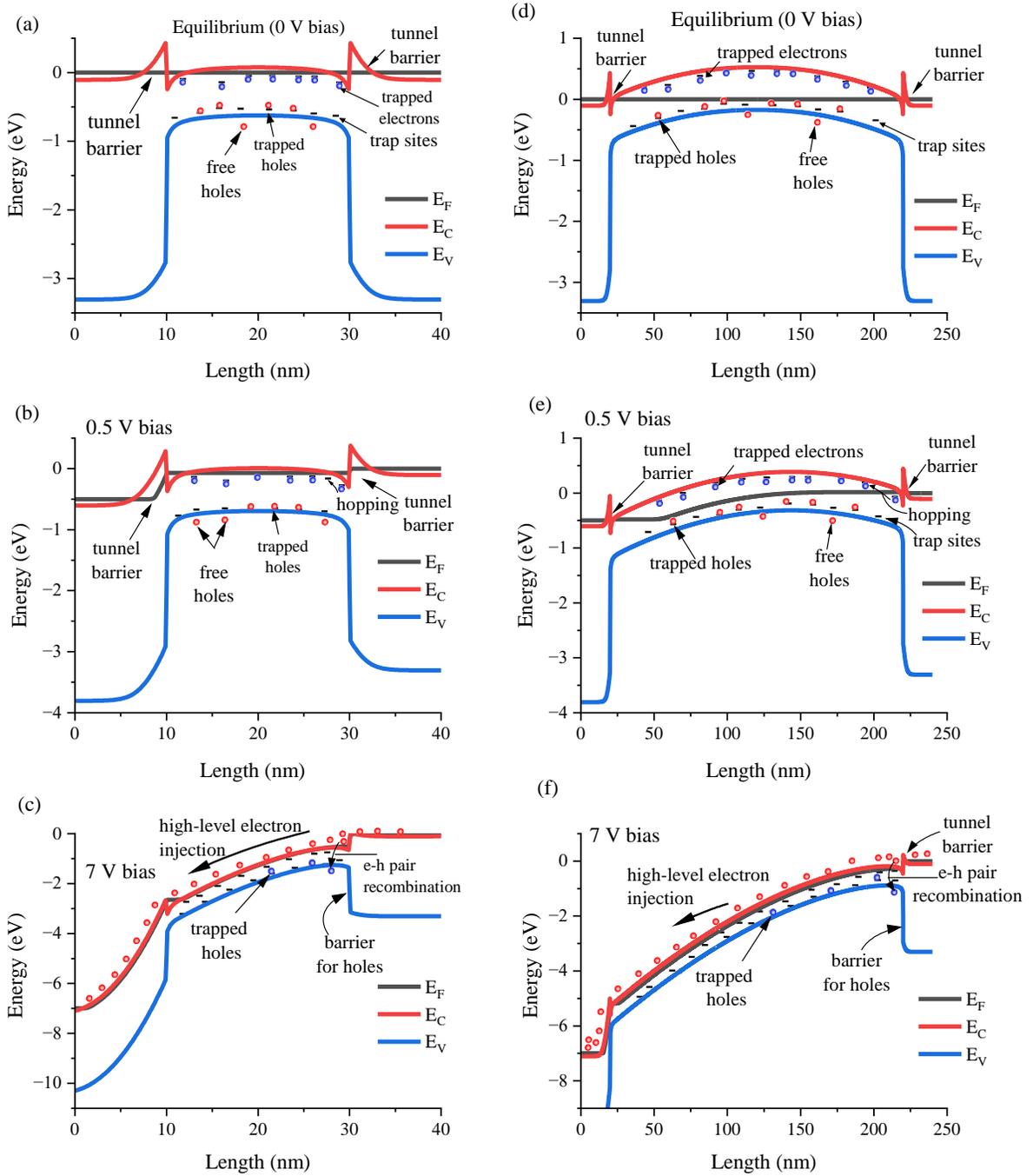

Fig. 4. Room temperature band diagrams of TiN / a-GST / TiN structure with two different GST length (distance between two TiN contacts) simulated using COMSOL Multiphysics semiconductor module [43]. (a) equilibrium case at zero bias for 20nm GST length, (b) under a mild bias of 0.5 V, (c) under high bias of 7 V, (e-f): equilibrium, 0.5V, and 5V cases respectively for 200nm GST length. In all cases, the de-trapped holes cannot escape the potential well and are likely to recombine with injected electrons.

$$I_{low\ field}(V,T) = I_{forward} - I_{reverse}$$

$$I_{low\ field}(V,T) = I_0 e^{-\frac{E_A - b\omega V k_B T}{k_B T}} - I_0 e^{-\frac{E_A - (b-1)\omega V k_B T}{k_B T}} \quad (5)$$

Here, $I_0$ is a current scaling factor, $b = d_{peak}/d_{trap}$ (Fig. 5) and $\omega = qd_{trap}/k_B TL$ are fitting parameters extracted from fitting equation (5) to low-field experimental data in Fig. 2. L is the amorphized length. As we increase voltage, the energy barrier for forward transmission decreases by $b\omega V k_B T$ and the reverse transmission energy barrier increases by $(b-1)\omega V k_B T$. For high enough bias, the barrier for forward transmission can be brought below the average kinetic energy of the electrons ($3k_B T/2$) in the source reservoir at the junction, leading to high level injection of electrons (Fig. 4c,f). However, holes in the a-GST region cannot escape even with a substantial bias due to the high potential barrier and are likely to recombine with electrons near the right interface. Since the charge exchanges at the interfaces and the resulting resistance drift in time increase the complexity of the study, we assume that the cell was recently reset and it is still in its metastable state in this study. Hence, we use the low-field transport model for stable a-GST (Fig. 2), which has to be modified for metastable a-GST.

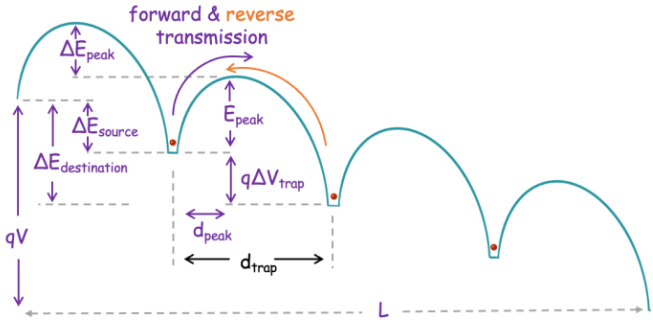

Fig. 5. A schematic illustration of conduction-band edge and n-type charge transport in a material with traps. The peak height of the energy barrier ($E_{peak}$) is modulated ($\Delta E_{peak}$) by the applied bias (V) for a crested barrier. The modulation in the barrier height depends on the ratio of the peak location ($d_{peak}$) and trap separation ($d_{trap}$).

Metastable a-GST was characterized using small voltage signals (0 to 0.2 V). Equation (5) can be approximated using the Taylor series expansion for small $V$ as:

$$I_{low\,field}(V,T) = I_0 e^{-\frac{E_A}{k_B T}}[e^{b\omega V} - e^{(b-1)\omega V}]$$

$$= I_0 e^{-\frac{E_A}{k_B T}}\left[\left\{1 + b\omega V + \frac{(b\omega V)^2}{2} + \cdots\right\}\right.$$
$$\left. -\left\{1 + (b-1)\omega V + \frac{((b-1)\omega V)^2}{2} + \cdots\right\}\right]$$

$$= I_0 e^{-\frac{E_A}{k_B T}}\left[\omega V + \frac{(b\omega V)^2}{2} - \frac{((b-1)\omega V)^2}{2}\right] \quad (6)$$

Since $b \approx 0.5$ for T > 300 K, extracted from the experiments[44], $b \approx 1 - b$. Therefore, the higher order terms in equation (6) cancel out and equation (6) simplifies to:

$$I_{low\,field}(V,T) = I_0 e^{-\frac{E_A}{k_B T}} \omega V \quad (7)$$

Which can be written in terms of current density and equated to drift current density, $J_{drift} = (1/\rho) E$, where $\rho$ is electrical resistivity and $E$ is electric field:

$$J_{low\,field}(V,T) = J_0 e^{-\frac{E_A}{k_B T}} \omega V = J_0 e^{-\frac{E_A}{k_B T}} \omega L E = \frac{1}{\rho} E \quad (8)$$

Here, $L$ is the length of the GST line cell and resistivity can be expressed as $\rho = e^{\frac{E_A}{k_B T}} \omega L / J_0$. Inserting the $\rho$ expression in equation (2), we calculate the value of $J_0$ as $\omega L / \rho_0$ = 188 MA/cm$^2$. Using this value for the $J_0$ pre-factor, the electric-field and temperature dependent current density model of metastable a-GST can be written as:

$$\boxed{J(V,T) = J_0 e^{-\frac{E_A}{k_B T}}[e^{b\omega V} - e^{(b-1)\omega V}]} \quad (9)$$

Electrostatic force on the electrons leading to $J_{drift}$ can be more accurately expressed in terms of the gradient of the conduction-band edge [57], which can be expressed in terms of the Peltier coefficient ($\Pi$):

$$F = -qE = -\nabla E_c = -\nabla(-q\Pi - \frac{3}{2}k_B T + E_f) \quad (10)$$

Here, $E_f$ is the Fermi energy and the external applied electric-field is expressed as $-\nabla V = -\nabla(E_f)$. Peltier coefficient, which can also be expressed as product of Seebeck coefficient[58] and temperature (ST), is the sum of average kinetic energy ($3k_B T/2$) and chemical potential energy ($E_c - E_f$) for electrons at a given location.

The total current has drift and diffusion components:

$$J_{total} = J_{drift} + J_{diffusion} \quad (11)$$

Using equation (10) drift current density can be written as:

$$J_{drift} = \frac{E}{\rho} = \frac{\nabla E_c}{\rho} = \frac{1}{\rho}\nabla(-q\Pi - \frac{3}{2}k_B T + E_f) \quad (12)$$

The diffusion current ($J_{diffusion}$) can be calculated using the open-circuit condition ($J_{total} = 0$):

$$J_{diffusion} = -J_{drift} = -\frac{1}{q\rho}\nabla\{-q\Pi - \frac{3}{2}k_B T + E_{f0}\} \quad (13)$$

Here, $E_{f0}(T)$ is the equilibrium Fermi level in a-GST. Hence,

$$J_{total} = \frac{1}{q\rho}\nabla(E_f - E_{f0}) = -\frac{1}{\rho}\nabla(V + \frac{E_{f0}}{q}) \quad (14)$$

Making the substitution for $V$ in equation (9) we can account for both drift and diffusion components, which come into play due to thermal gradients:

$$J(V,T) = J_0 e^{-\frac{E_A}{k_B T}}\left[e^{b\omega(V+\frac{L}{q}\nabla E_{f0})} - e^{(b-1)\omega(V+\frac{L}{q}\nabla E_{f0})}\right] \quad (15)$$

We do not have a measured $E_{f0}(T)$. Hence, we approximate the function based on the 300 K and T$_{metal}$ values. At 300 K, the bandgap of a-GST is 0.7 eV, and the Fermi level is estimated to be 0.15 eV above the valance band edge, $E_{f0}(300K) = E_i - 0.2\,eV$, where $E_i$ corresponds to the middle of the bandgap. At metal transition temperature ($T_{metal} \approx 930\,K$), the bandgap collapses to zero, thus $E_{f0}(930K) = 0\,eV$. Assuming $E_{f0}(T)$ to be a linear function of T, we can estimate $dE_{f0}(T)/dT$ to be 0.32 meV/K.

The barrier height for forward transmission can be expressed as $E_{peak} = E_A - b\omega(V + L/q \cdot \nabla E_{f0})k_B T$, and $E_{peak} \geq 0$. Hence, under extreme bias ($V \geq E_A/b\omega k_B T - L/q \cdot \nabla E_{f0}$), $E_{peak} = 0$, and current density, $J$, converges to $J_0$. This corresponds to a saturation velocity ($\vartheta_{sat} = J_0/qn$) of ~$10^5$ cm/s, assuming carrier concentration $n = 10^{22}$ cm$^{-3}$[59]. As a comparison, $\vartheta_{sat}$ of amorphous Silicon is $1.7\times10^4$ cm/s[60] and for crystalline Silicon is $1\times10^7$ cm/s[61], [62].

## III. THERMAL TRANSPORT IN AMORPHOUS GST

Thermal transport and energy exchanges are modeled using:

$$\underbrace{dC_p \frac{dT}{dt}}_{\text{Heat absorbed by the material}} \underbrace{-\nabla \cdot (k\nabla T)}_{\text{Fourier conduction}} = \underbrace{-\nabla V \cdot J}_{\text{Joule heat}} \underbrace{-\nabla \cdot (JST)}_{\text{Thermo-electric heat}} + \underbrace{Q_H}_{\text{Latent heat of phase change}} \quad (16)$$

where $d$ is the mass density, $C_p$ is the specific heat, $t$ is the time, $k$ is the thermal conductivity ($k_{a-GST} = 0.27$ W/m·K), $Q_H$ is the latent heat of phase change. The thermoelectric terms introduce an asymmetry. Equations (15) and (16) are solved self-consistently in COMSOL to model the device dynamics.

## IV. FILAMENTARY CONDUCTION

The $E_A(T)$ extracted from metastable conductivity corresponds to an effective activation energy (Fig. 1b) due to a distribution of trap levels inside GST that contribute to conduction, fixed charges that change the local electrostatic potential and local compositional variations in the disordered material. Hence, we assume a local random distribution of activation energy, $E_A = 0.33\ eV \pm \sigma_E$, $\sigma_E = 0.05\ eV$ (15.6% variation), to model electronic transport in the disordered a-GST in 2D with W = 50 nm × L = 20 nm × $t_{out\text{-}of\text{-}plane}$ = 2 nm sandwiched between two TiN contacts (Fig. 6 -Fig. 9). We then vary these parameters to analyze their impact (Fig. 10).

We generate the local variations in $E_A$ in the finite-element simulations by starting with a random $E_A$ map with 2 nm x 2 nm regions of uniform $E_A$, resembling a quilt, and use a

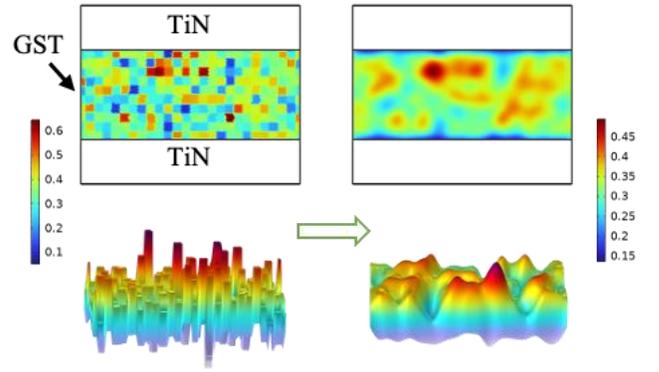

Fig. 6. Local distribution of $E_A$ in a-GST ($W_{GST} \times L_{GST}$ = 50 nm × 20 nm) sandwiched between two TiN contacts ($W_{TiN} \times L_{TiN}$ = 50 nm ×10 nm). The $E_A$ block size is 2 nm × 2 nm before smoothing (left), smoothed $E_A$ after diffusion (right) and 3D views of the distribution (bottom).

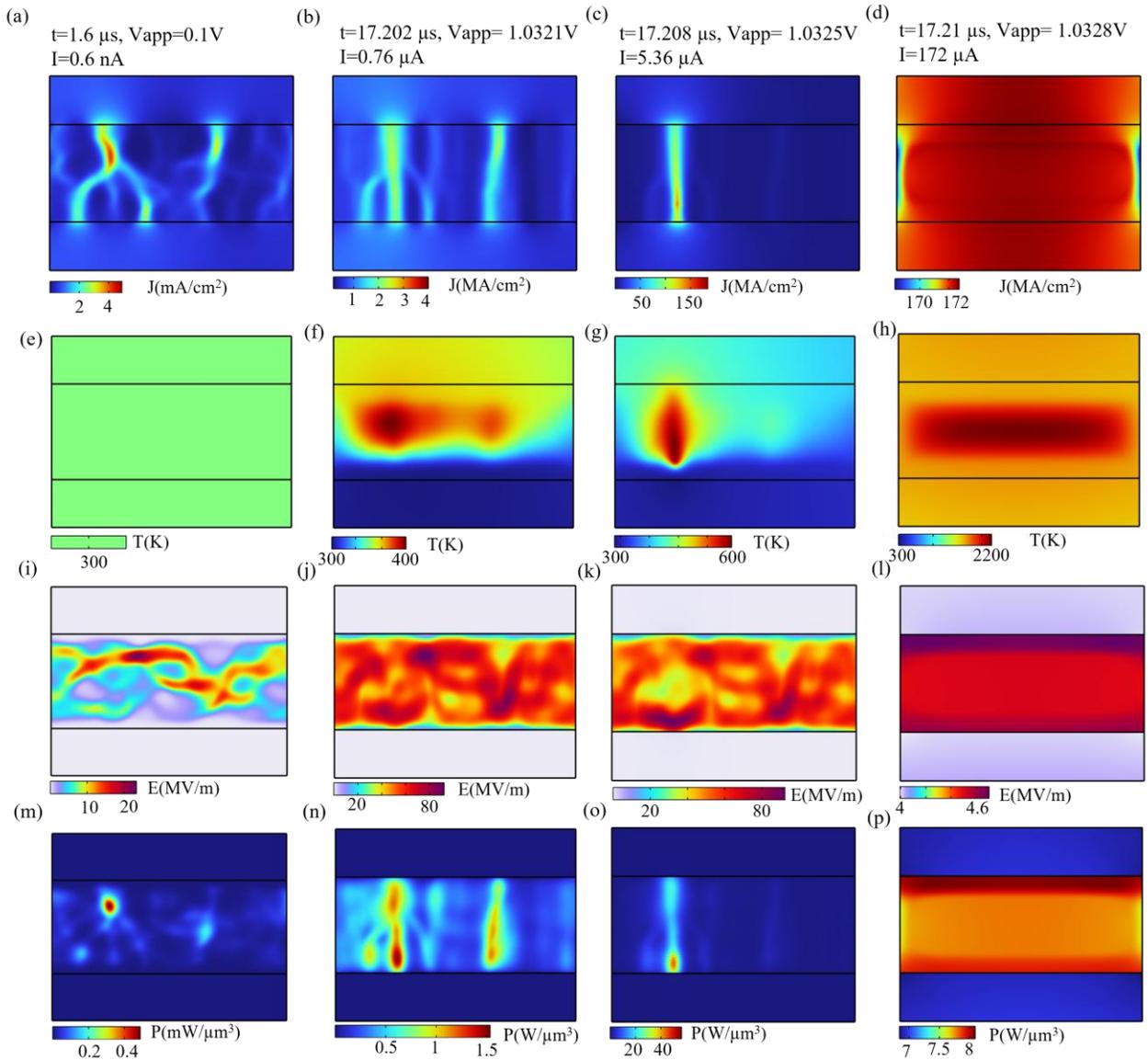

Fig. 7. (a-d): The current density profiles in 50 nm × 20 nm × 2 nm GST region with 1kΩ load resistor showing filamentation at different times during the 0 V to 3 V 50 µs voltage ramp for $T_{ambient}$ = 300 K, (e-h): Corresponding temperature profiles, (i-l): Electric field profile inside the device, (m-p): power density profile.

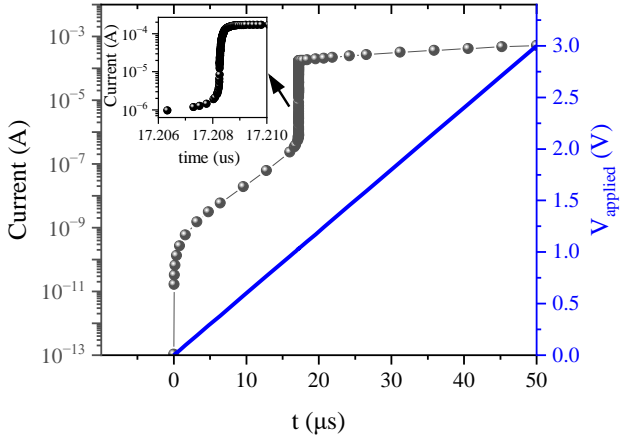

Fig. 8. Current and applied voltage versus time characteristics of the simulation results shown in Fig. 7. Thermal runaway and snap-back takes place in < 1 ns.

diffusion function to smooth $E_A$ at the mesh interfaces to model the local variations (Fig. 6):

$$\frac{\partial E_A}{\partial t} + \nabla \cdot (-c \nabla E_A) = 0 \quad (17)$$

This also allows us to have continuous (non-abrupt) derivatives across mesh-points that enables convergence of the numerical computations during the simulations. Longer diffusion times or higher diffusion constant, $c$, lead to smoother $E_A$ variations, and smaller $\sigma_E$ (Fig. 6, Fig. 10a).

In the device simulations, we apply a bias to the top TiN contact, and the bottom TiN contact is connected through a 1 kΩ load resistor to ground. We ramp the bias from 0 V to 3 V in 50 $\mu s$ and observe the current (Fig. 8), local current density, temperature, electric field and power density (Fig. 7, organized as 4 columns for different points in time, corresponding to different applied voltages). The filaments start forming at very low bias voltages (Fig. 7a). As we increase bias, the filaments start aligning with the electric field (Fig. 7b). Once we reach a critical bias ($V_{Snap}$), a single filament gets into thermal runaway in less than 1 ns and dominates current conduction (Fig. 7c). Beyond this point, the increase of voltage leads to melting and gradual widening of the molten filament until the whole GST region melts (Fig. 7e,f,g). The corresponding I-V characteristics show a snapback behavior that is commonly observed in experiments (Fig. 9). The snapback electric field ($E_{Snap}$) decreases with ambient temperature ($T_{ambient}$) (Fig. 10b). The current at which the snapback is initiated, snapback current ($I_{Snap} \approx 0.6\ \mu A$ for 300 K), increases with $T_{ambient}$ (Fig. 9a - inset). At the snapback condition, a single filament of ~ 2 nm width forms for $T_{ambient} \leq 400\ K$ (Fig. 10c). The filaments are not well defined for $T_{ambient} > 400\ K$. After the snapback, voltage across the GST ($V_{GST}$) keeps decreasing as the filament keeps widening, until the whole device melts. When the resistance of GST and load resistor become equal, the maximum power point is reached[63]. The high-resistance (before snapback) and the low-resistance (after snapback) states of the device show 0.2 G$\Omega$ at 0.1 V and 1.75 k$\Omega$ at 0.25V respectively for $T_{ambient} = 300\ K$.

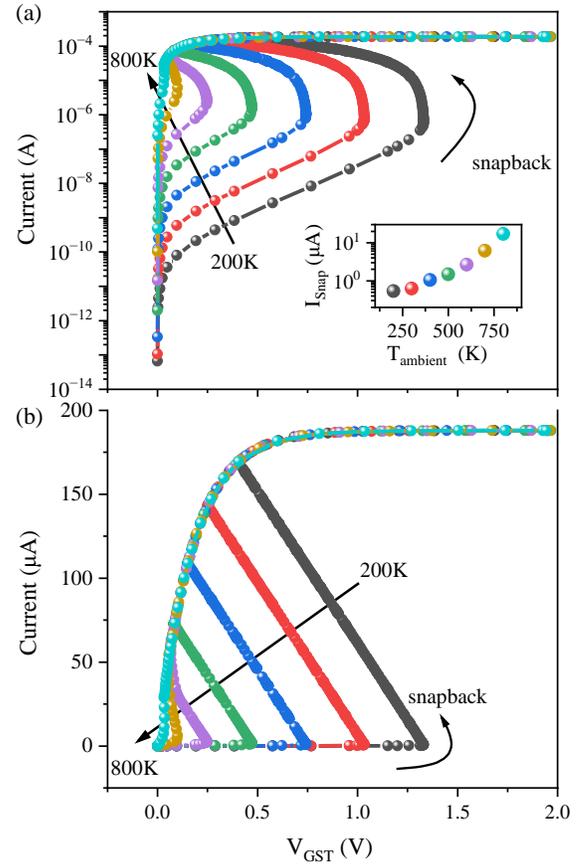

Fig. 9. I-V characteristics of the sandwich structure with 50 nm × 20 nm × 2 nm GST region using the filamentary conduction model for $T_{ambient}$ ranging from 200 K to 800 K in 100 K steps.

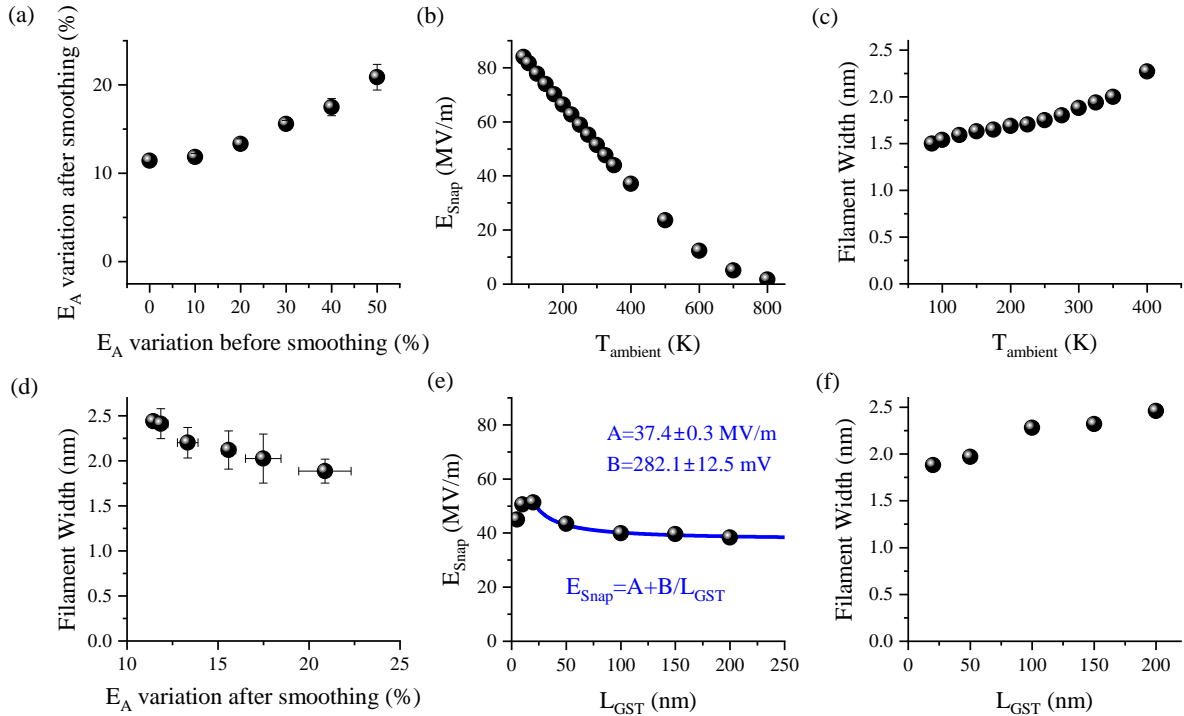

Fig. 10. Effect of variations in model parameters in a PCM cell with 50 nm × 20 nm × 2 nm GST region. (a-d): Correlation of narrowest single filament width with $E_A$ standard deviation, ambient temperature, and length of GST (distance between two contacts). (e-f): Correlation of snapback electric filed with GST length and ambient temperature.

We simulated devices by varying $T_{ambient}$, $\sigma_E$, mesh sizes, random number seeds (distributions) for $E_A$ and $L_{GST}$, and observed the impact on filament width and snapback electric field ($E_{Snap}$) (Fig. 10). We estimate the width of narrowest single filament formed in a device by calculating the full width at half of maximum current density. Mesh size, different random number seeds, load resistor values or ramp-rate of the applied bias does not result in significant changes in filament width. Lower $T_{ambient}$ lead to slight narrower (~ 1.5 nm) filaments (Fig. 10c). Higher $\sigma_E$ results in narrower filaments (Fig. 10d). However, the filament width (ranging from ~2.45 nm to ~1.65 nm) is not very sensitive to $\sigma_E$ and the device-to-device variation in filament width is < 15 % (Fig. 10d). Even if there is no variation in $E_A$ within the bulk of GST and at the interfaces, the numerical noise leads to filament formation due to the positive-feedback in the process of thermal runaway (the first data points in Fig. 10a,d). This suggests that even perfectly uniform structures with insulating layers are expected experience filamentary conduction (due to thermal excitations), thermal runaway and dielectric breakdown.

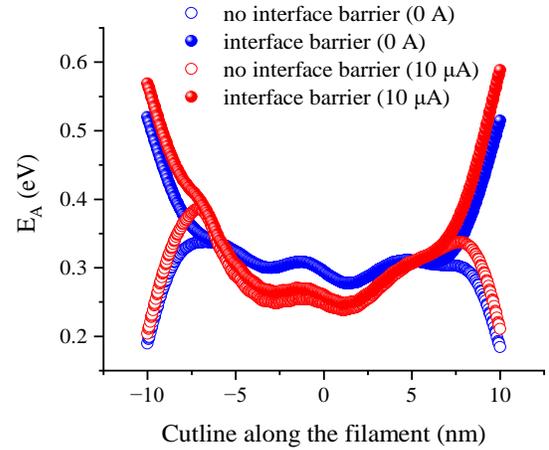

Fig. 11. Activation energy profile along the filament for low and high interface $E_A$ at GST/TiN interface.

Higher $L_{GST}$ results in reduced $E_{Snap}$, saturating at ~38 MV/m for $L_{GST} > 200$ nm (Fig. 10e), where the interface effects are not significant. $E_{Snap}$ increases to ~ 50 MV/m if $L_{GST}$ is reduced below 50 nm. This is due to increased thermal coupling of the filament to the contacts, requiring higher electric fields to initiate thermal runaway in the filament (Fig. 10f). We observe a decline in $E_{Snap}$ for $L_{GST} < 20$ nm, which is due to dominance of the low $E_A$ used at the GST/TiN interfaces in the model and the increased probability of low $E_A$ percolation paths between the two contacts.

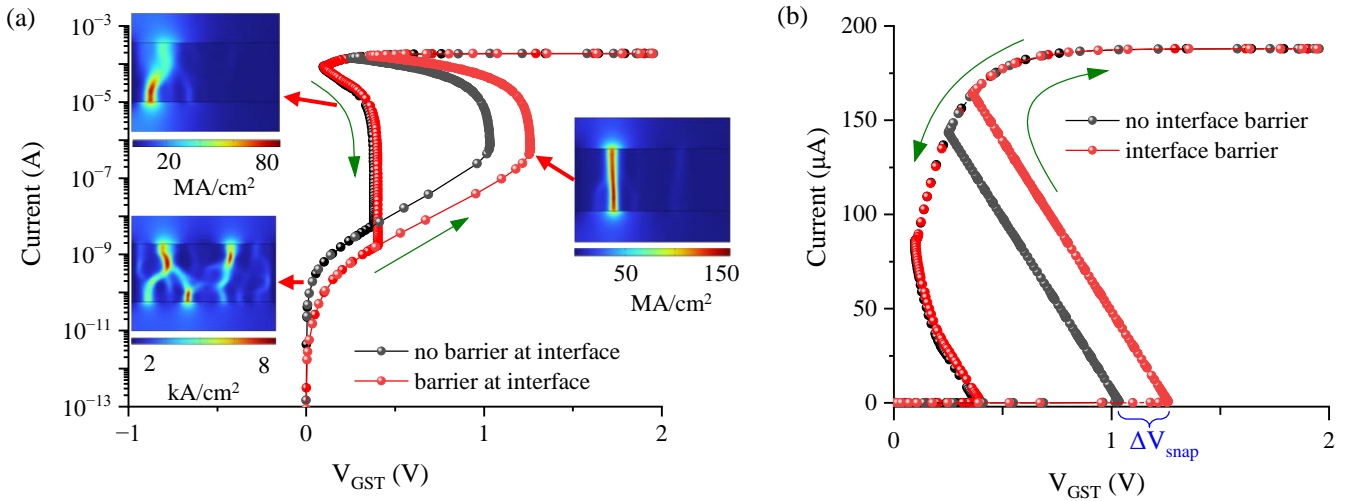

Fig. 12. Room temperature I-V characteristics of the sandwich structure with 50 nm × 20 nm × 2 nm GST region for low and high interface $E_A$ at GST/TiN interface during a 0 - 3V triangular voltage pulse with 50 $\mu s$ rise and fall times in log (a) and linear scales (b). Insets show the current density distribution in the device at the indicated points in the voltage ramp up and down.

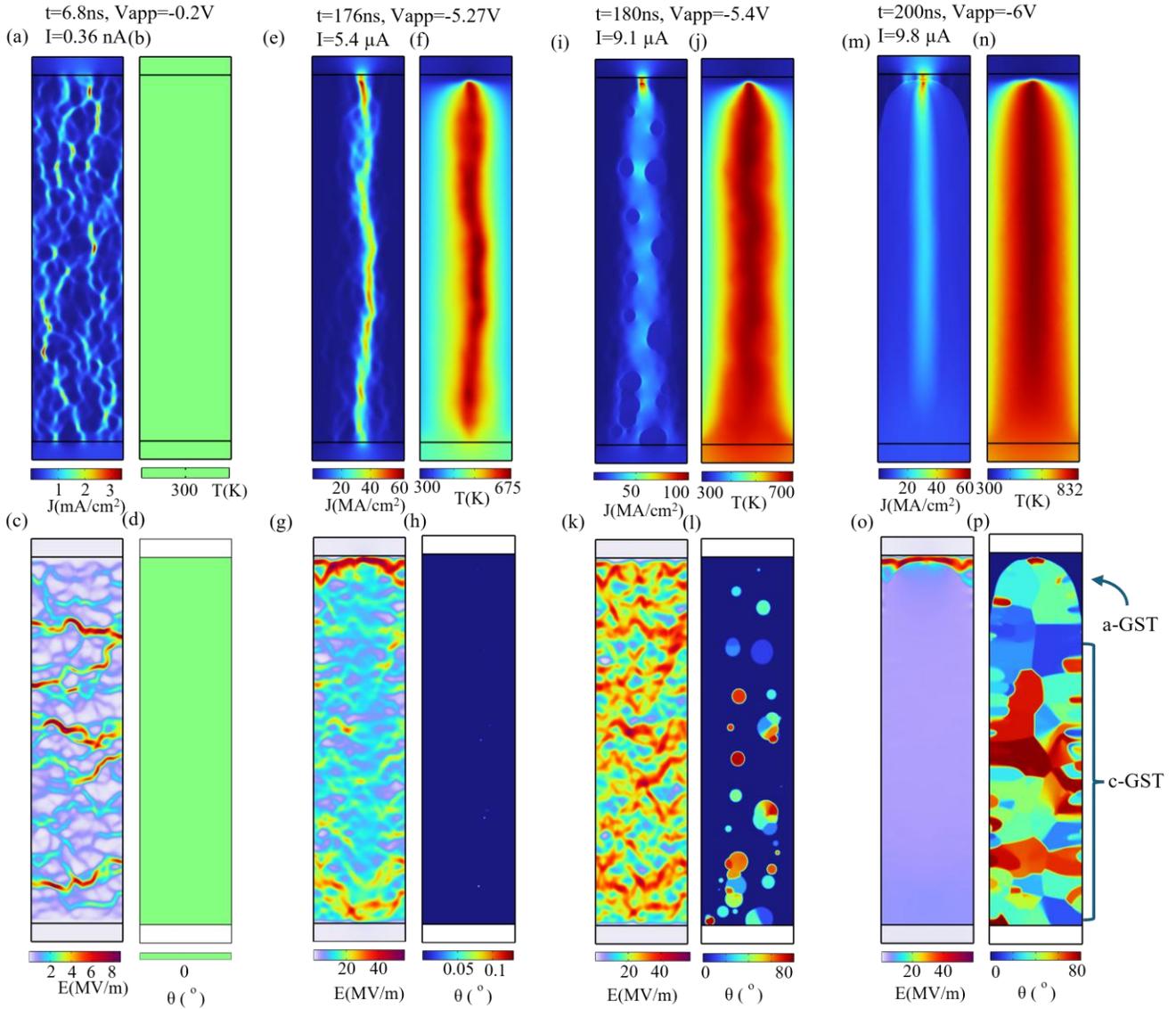

Fig. 13. Set operation in a long device with 50 nm × 200 nm × 2 nm GST region with 100 kΩ load resistor. (a, e, i, m): Current density at different times during the 0 V to -6 V 200 ns voltage ramp, (b, f, j, n): corresponding temperature profiles, (c, g, k, o): electric field profiles, (d, h, l, p): orientation angle of the grains.

Fabrication processes and surface preparation are expected to change the chemical composition and hence the $E_A$ at the GST/TiN interfaces. We compared the low interface $E_A$ case we analyzed (Fig. 6 - Fig. 10) to high interface $E_A$ cases for $L_{GST}$ = 20 nm (Fig. 11) where a triangular pulse with 50 $\mu$s rise and fall times are used. High interface $E_A$ is expected if an energy barrier is formed at the GST/TiN interfaces, and results in a higher $V_{Snap}$ (Fig. 12) but $I_{Snap}$ does not change. The behavior of the two devices is the same on the low-resistance state (after snap-back and ramp down). We observe that the filaments are no longer aligned with the electric field on the ramp-down, which appears to be due to increased local temperature around the filaments (Fig. 12a insets). Filaments with high current density can be maintained with ~ 80 $\mu$A and $V_{GST}$ = ~ 0.25 V. Hence, set operation can be achieved with reduced power by reducing the applied voltage after initiating a molten filament (breakdown).

In the case of the longer devices (~200 nm), the percolation paths are clearly visible at the earlier stages (Fig. 13a-h) and the current still collapses onto a single filament (Fig. 13i-l). In this example (Fig. 13), using a 100 kΩ load resistor and a voltage at the top contact ramped from 0V to -6V in 200 ns, we observe full crystallization (set) of a long PCM device. Nucleation starts mainly at the areas around the filament which is visible in the grain orientation angle plot (Fig. 13l) and also in the current density plot (Fig. 13i). Here, the phase-change physics, described in our earlier publications[38]–[41], is solved self-consistently with equations (15) and (16) to capture the nucleation, growth and amorphization dynamics. The top-to-bottom asymmetry in thermal profile is due to thermoelectric effects. As the load resistor is quite large, the filaments do not reach $T_{melt}$.

In order to understand the impact of the expected local variations on the experimentally observed (effective) $E_A$, we fitted the I-V curves extracted from the simulations (Fig. 14) to the current density model of equation (9) up to the snapback voltage and extracted the effective $E_A$ as we do in the experimental studies (Fig. 1). The effective $E_A$ we calculate is very close to the average $E_A$ we use in the simulations (Fig. 14). Hence, the local variations in $E_A$ do not appear to be a concern for calculating $E_A$ from experimental I-V characteristics.

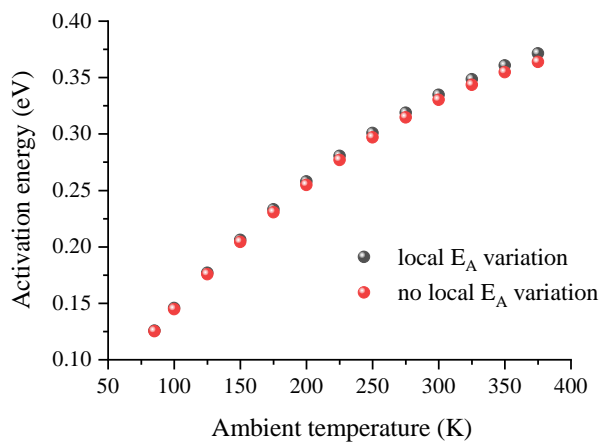

Fig. 14. Effective $E_A$ extracted from I-V characteristics obtained from the simulations (●) and $E_A$ from equation (4), where there is no local $E_A$ variation (●).

## CONCLUSION

Local variations in conductivity lead to filamentary conduction in disordered materials such as amorphous phase change and ovonic threshold switch (OTS)[64]–[67] materials. Our computational analysis shows formation of a single filament with ~ 2 nm diameter in each device, which gets into thermal runaway below 1 $\mu A$ current for $T_{ambient} \leq 400\ K$, giving rise to the snapback behavior that has been experimentally observed. Hence, it is possible to thermally switch the PCM as well as the OTS devices from the high-resistance state to low-resistance state and low switching current (~1 $\mu A$) is not a clear identifier of electronic switching. However, the transition we observe in our experimental studies, showing a low-field to high-field transition (E ~ 20 MV/m) in the order of 10 pA around 150 K for relatively long (200 nm – 700 nm amorphized lengths) stable cells, suggest a change in the electronic processes that limit current[44], [59]. The device-scale potential barrier that forms for the electrons in TiN / a-GST / TiN structures, due to the charge exchanges at the interfaces, scale down for shorter devices (Fig. 4a vs. Fig. 4d). However, the potential well that forms for the holes are very deep. Hence, excess holes stored in the amorphous region can only be annihilated by recombination with injection of electrons into a-GST. We included the thermoelectric effects and the contribution of the diffusion current in our simulations. However, these effects do not have a major impact on filament formation. The insensitivity of the filament width to device dimensions and activation energy variations, and the consistent behavior of filamentary conduction, thermal runaway and snap-back behavior suggests that conduction in highly resistive materials is filamentary regardless of their level of disorder. This work opens the scope of modeling nanoscale variations of material properties in various ordered (uniform) and disordered material systems and devices. We plan to extend this model to capture dynamic changes in $E_A$ due to local electric field, current density and temperature, which may capture resistance drift in PCM and formation of conductive filaments in PCM and other resistive random access memory (RRAM)[68], [69] devices as well as reverse biased pn junction and p-i-n diodes[70], magnetic tunnel junctions[71] and Josephson junctions[72].

## ACKNOWLEDGEMENTS

The devices were fabricated at IBM Watson Research Center under a Joint Study Agreement.

## REFERENCES

[1] M. Gu, Q. Zhang, and S. Lamon, "Nanomaterials for optical data storage," *Nature Reviews Materials 2016 1:12*, vol. 1, no. 12, pp. 1–14, Oct. 2016, doi: 10.1038/natrevmats.2016.70.

[2] W. Zhang, R. Mazzarello, M. Wuttig, and E. Ma, "Designing crystallization in phase-change materials for universal memory and neuro-inspired computing," *Nat Rev Mater*, vol. 4, no. 3, pp. 150–168, Mar. 2019, doi: 10.1038/S41578-018-0076-X.

[3] "Does AI have a hardware problem?," *Nature Electronics 2018 1:4*, vol. 1, no. 4, pp. 205–205, Apr. 2018, doi: 10.1038/s41928-018-0068-2.

[4] "Big data needs a hardware revolution," *Nature*, vol. 554, no. 7691, pp. 145–146, Feb. 2018, doi: 10.1038/D41586-018-01683-1.

[5] M. K. Qureshi, V. Srinivasan, and J. A. Rivers, "Scalable high performance main memory system using phase-change memory technology," *Proc Int Symp Comput Archit*, pp. 24–33, 2009, doi: 10.1145/1555754.1555760.

[6] G. C. Han, J. J. Qiu, L. Wang, W. K. Yeo, and C. C. Wang, "Perspectives of read head technology for 10 Tb/in2 recording," *IEEE Trans Magn*, vol. 46, no. 3 PART 1, pp. 709–714, 2010, doi: 10.1109/TMAG.2009.2034866.

[7] H. S. P. Wong and S. Salahuddin, "Memory leads the way to better computing," *Nature Nanotechnology 2015 10:3*, vol. 10, no. 3, pp. 191–194, Mar. 2015, doi: 10.1038/nnano.2015.29.

[8] M. Wuttig and N. Yamada, "Phase-change materials for rewriteable data storage," *Nature Materials 2007 6:11*, vol. 6, no. 11, pp. 824–832, 2007, doi: 10.1038/nmat2009.

[9] R. Waser and M. Aono, "Nanoionics-based resistive switching memories," *Nature Materials 2007 6:11*, vol. 6, no. 11, pp. 833–840, Nov. 2007, doi: 10.1038/nmat2023.

[10] A. D. Kent and D. C. Worledge, "A new spin on magnetic memories," *Nat Nanotechnol*, vol. 10, no. 3, pp. 187–191, Mar. 2015, doi: 10.1038/NNANO.2015.24.

[11] J. F. Scott and C. A. Paz De Araujo, "Ferroelectric Memories," *Science (1979)*, vol. 246, no. 4936, pp. 1400–1405, Dec. 1989, doi: 10.1126/SCIENCE.246.4936.1400.

[12] F. Pan, S. Gao, C. Chen, C. Song, and F. Zeng, "Recent progress in resistive random access memories: Materials, switching mechanisms, and performance," *Materials Science and Engineering: R: Reports*, vol. 83, no. 1, pp. 1–59, Sep. 2014, doi: 10.1016/J.MSER.2014.06.002.

[13] S. W. Fong, C. M. Neumann, and H. S. P. Wong, "Phase-Change Memory - Towards a Storage-Class Memory," *IEEE Trans Electron Devices*, vol. 64, no. 11, pp. 4374–4385, Nov. 2017, doi: 10.1109/TED.2017.2746342.

[14] G. W. Burr, B. N. Kurdi, J. C. Scott, C. H. Lam, K. Gopalakrishnan, and R. S. Shenoy, "Overview of candidate device technologies for storage-class memory," *IBM J Res Dev*, vol. 52, no. 4–5, pp. 449–464, 2008, doi: 10.1147/RD.524.0449.

[15] M. Le Gallo and A. Sebastian, "An overview of phase-change memory device physics," *J Phys D Appl Phys*, vol. 53, no. 21, p. 213002, May 2020, doi: 10.1088/1361-6463/ab7794.

[16] S. Raoux, F. Xiong, M. Wuttig, and E. Pop, "Phase change materials and phase change memory," *MRS Bull*, vol. 39, no. 8, pp. 703–710, Aug. 2014, doi: 10.1557/MRS.2014.139.

[17] G. W. Burr, M. J. Breitwisch, M. Franceschini, D. Garetto, K. Gopalakrishnan, B. Jackson, B. Kurdi, C. Lam, L. A. Lastras, A. Padilla, B. Rajendran, S. Raoux, and R. S. Shenoy, "Phase change memory technology," *Journal of Vacuum Science and Technology B*, vol. 28, no. 2, pp. 223–262, Mar. 2010, doi: 10.1116/1.3301579.

[18] L. Wang, L. Tu, and J. Wen, "Application of phase-change materials in memory taxonomy," *Sci Technol Adv Mater*, vol. 18, no. 1, pp. 406–429, Jan. 2017, doi: 10.1080/14686996.2017.1332455.

[19] M. Chen, K. A. Rubin, and R. W. Barton, "Compound materials for reversible, phase-change optical data storage," *Appl Phys Lett*, vol. 49, no. 9, pp. 502–504, Dec. 1986, doi: 10.1063/1.97617.

[20] N. Yamada, E. Ohno, N. Akahira, K. Nishiuchi, K. Nagata, and M. Takao, "High speed overwritable phase change optical disk material," *Jpn J Appl Phys*, vol. 26, no. S4, pp. 61–66, Jan. 1987, doi: 10.7567/JJAPS.26S4.61/XML.

[21] N. Yamada, E. Ohno, K. Nishiuchi, N. Akahira, and M. Takao, "Rapid-phase transitions of GeTe-Sb2Te3 pseudobinary amorphous thin films for an optical disk memory," *J Appl Phys*, vol. 69, no. 5, pp. 2849–2856, Mar. 1991, doi: 10.1063/1.348620.

[22] J. L. F. Da Silva, A. Walsh, and H. Lee, "Insights into the Structure of the Stable and Metastable (GeTe)m(Sb2Te3)n Compounds," *Phys Rev B Condens Matter Mater Phys*, vol. 78, no. 22, 2008, Jan. 2008, doi: 10.1103/PHYSREVB.78.224111.

[23] D. Ielmini, "Electrical Transport in Crystalline and Amorphous Chalcogenide," *Phase Change Memory: Device Physics, Reliability and Applications*, pp. 11–39, Nov. 2018, doi: 10.1007/978-3-319-69053-7_2.

[24] V. Sousa and G. Navarro, "Material Engineering for PCM Device Optimization," *Phase Change Memory: Device Physics, Reliability and Applications*, pp. 181–222, Nov. 2018, doi: 10.1007/978-3-319-69053-7_7.

[25] K. Cil, F. Dirisaglik, L. Adnane, M. Wennberg, A. King, A. Faraclas, M. B. Akbulut, Y. Zhu, C. Lam, A. Gokirmak, and H. Silva, "Electrical resistivity of liquid Ge2Sb2Te5 based on thin-film and nanoscale device measurements," *IEEE Trans Electron Devices*, vol. 60, no. 1, pp. 433–437, 2013, doi: 10.1109/TED.2012.2228273.

[26] M. Xu, Y. Q. Cheng, L. Wang, H. W. Sheng, Y. Meng, W. G. Yang, X. D. Han, and E. Ma, "Pressure tunes electrical resistivity by four orders of magnitude in amorphous Ge2Sb2Te5 phase-change memory alloy," *Proc Natl Acad Sci U S A*, vol. 109, no. 18, May 2012, doi: 10.1073/PNAS.1119754109.

[27] Ohshima and Norikazu, "Crystallization of germanium-antimony-tellurium amorphous thin film sandwiched between various dielectric protective films," *JAP*, vol. 79, no. 11, pp. 8357–8363, Jun. 1996, doi: 10.1063/1.362548.

[28] H. K. Peng, K. Cil, A. Gokirmak, G. Bakan, Y. Zhu, C. S. Lai, C. H. Lam, and H. Silva, "Thickness dependence of the amorphous-cubic and cubic-hexagonal phase transition temperatures of GeSbTe thin films on silicon nitride," *Thin Solid Films*, vol. 520, no. 7, pp. 2976–2978, Jan. 2012, doi: 10.1016/J.TSF.2011.11.033.

[29] A. Faraclas, G. Bakan, L. Adnane, F. Dirisaglik, N. E. Williams, A. Gokirmak, and H. Silva, "Modeling of thermoelectric effects in phase change memory cells," *IEEE Trans Electron Devices*, vol. 61, no. 2, pp. 372–378, 2014, doi: 10.1109/TED.2013.2296305.

[30] T. Siegrist, P. Jost, H. Volker, M. Woda, P. Merkelbach, C. Schlockermann, and M. Wuttig, "Disorder-induced localization in crystalline phase-change materials," *Nat Mater*, vol. 10, no. 3, pp. 202–208, 2011, doi: 10.1038/NMAT2934.

[31] L. Adnane, F. Dirisaglik, A. Cywar, K. Cil, Y. Zhu, C. Lam, A. F. M. Anwar, A. Gokirmak, and H. Silva, "High temperature electrical resistivity and Seebeck coefficient of Ge2Sb2Te5 thin films," *J Appl Phys*, vol. 122, no. 12, Sep. 2017, doi: 10.1063/1.4996218.

[32] F. Dirisaglik, G. Bakan, Z. Jurado, S. Muneer, M. Akbulut, J. Rarey, L. Sullivan, M. Wennberg, A. King, L. Zhang, R. Nowak, C. Lam, H. Silva, and A. Gokirmak, "High speed, high temperature electrical characterization of phase change materials: metastable phases, crystallization dynamics, and resistance drift," *Nanoscale*, vol. 7, no. 40, pp. 16625–16630, Oct. 2015, doi: 10.1039/C5NR05512A.

[33] A. Redaelli, "Phase change memory: Device physics, reliability and applications," *Phase Change Memory:*

Accepted in PSS Rapid Research Letters for Special Issue on phase change memory (EPCOS 2024)    11Device Physics, Reliability and Applications, pp. 1–330, Nov. 2017, doi: 10.1007/978-3-319-69053-7.

[34] S. R. Ovshinsky, "An introduction to ovonic research," *J Non Cryst Solids*, vol. 2, no. C, pp. 99–106, Jan. 1970, doi: 10.1016/0022-3093(70)90125-0.

[35] A. Faraclas, N. Williams, A. Gokirmak, and H. Silva, "Modeling of set and reset operations of phase-change memory cells," *IEEE Electron Device Letters*, vol. 32, no. 12, pp. 1737–1739, Dec. 2011, doi: 10.1109/LED.2011.2168374.

[36] A. Faraclas, N. Williams, F. Dirisaglik, K. Cil, A. Gokirmak, and H. Silva, "Operation dynamics in phase-change memory cells and the role of access devices," *Proceedings - 2012 IEEE Computer Society Annual Symposium on VLSI, ISVLSI 2012*, pp. 78–83, 2012, doi: 10.1109/ISVLSI.2012.48.

[37] G. Bakan, B. Gerislioglu, F. Dirisaglik, Z. Jurado, L. Sullivan, A. Dana, C. Lam, A. Gokirmak, and H. Silva, "Extracting the temperature distribution on a phase-change memory cell during crystallization," *J Appl Phys*, vol. 120, no. 16, p. 164504, Oct. 2016, doi: 10.1063/1.4966168/13290383/164504_1_ACCEPTED_MANUSCRIPT.PDF.

[38] J. Scoggin, Z. Woods, H. Silva, and A. Gokirmak, "Modeling heterogeneous melting in phase change memory devices," *Appl Phys Lett*, vol. 114, no. 4, Jan. 2019, doi: 10.1063/1.5067397/1022986.

[39] J. Scoggin, R. S. Khan, H. Silva, and A. Gokirmak, "Modeling and impacts of the latent heat of phase change and specific heat for phase change materials," *Appl Phys Lett*, vol. 112, no. 19, May 2018, doi: 10.1063/1.5025331/14143006/193502_1_ONLINE.PDF.

[40] Z. Woods, J. Scoggin, A. Cywar, L. Adnane, and A. Gokirmak, "Modeling of Phase-Change Memory: Nucleation, Growth, and Amorphization Dynamics during Set and Reset: Part II - Discrete Grains," *IEEE Trans Electron Devices*, vol. 64, no. 11, pp. 4472–4478, Nov. 2017, doi: 10.1109/TED.2017.2745500.

[41] Z. Woods and A. Gokirmak, "Modeling of Phase-Change Memory: Nucleation, Growth, and Amorphization Dynamics during Set and Reset: Part I- Effective Media Approximation," *IEEE Trans Electron Devices*, vol. 64, no. 11, pp. 4466–4471, Nov. 2017, doi: 10.1109/TED.2017.2745506.

[42] G. Bakan, N. Khan, H. Silva, and A. Gokirmak, "Higherature thermoelectric transport at small scales: Thermal generation, transport and recombination of minority carriers," *Sci Rep*, vol. 3, 2013, doi: 10.1038/SREP02724.

[43] "COMSOL 6.2 - COMSOL Multiphysics Reference Manual." https://doc.comsol.com/6.2/doc/com.comsol.help.comsol/html_COMSOL_ReferenceManual.html (accessed Nov. 29, 2024).

[44] R. S. Khan, A. H. Talukder, F. Dirisaglik, H. Silva, and A. Gokirmak, "Accelerating and Stopping Resistance Drift in Phase Change Memory Cells via High Electric Field Stress," *arXiv preprint arXiv:2002.12487*, 2020.

[45] R. S. Khan, A. B. M. H. Talukder, F. Dirisaglik, A. Gokirmak, and H. Silva, "Stopping Resistance Drift in Phase Change Memory Cells," in *2020 Device Research Conference (DRC)*, 2020, pp. 1–2.

[46] S. Muneer, J. Scoggin, F. Dirisaglik, L. Adnane, A. Cywar, G. Bakan, K. Cil, C. Lam, H. Silva, and A. Gokirmak, "Activation energy of metastable amorphous Ge2Sb2Te5 from room temperature to melt," *AIP Adv*, vol. 8, no. 6, Jun. 2018, doi: 10.1063/1.5035085/311035.

[47] A. Talukder, M. Kashem, M. Hafiz, R. Khan, F. Dirisaglik, H. Silva, and A. Gokirmak, "Electronic transport in amorphous Ge2Sb2Te5 phase-change memory line cells and its response to photoexcitation," *Appl Phys Lett*, vol. 124, no. 26, Jun. 2024, doi: 10.1063/5.0196842.

[48] T. Kato and K. Tanaka, "Electronic properties of amorphous and crystalline Ge2Sb 2Te5 films," *Japanese Journal of Applied Physics, Part 1: Regular Papers and Short Notes and Review Papers*, vol. 44, no. 10, pp. 7340–7344, Oct. 2005, doi: 10.1143/JJAP.44.7340.

[49] B. S. Lee, J. R. Abelson, S. G. Bishop, D. H. Kang, B. K. Cheong, and K. B. Kim, "Investigation of the optical and electronic properties of Ge 2Sb 2Te 5 phase change material in its amorphous, cubic, and hexagonal phases," *J Appl Phys*, vol. 97, no. 9, May 2005, doi: 10.1063/1.1884248/914044.

[50] H. Tong, Z. Yang, N. N. Yu, L. J. Zhou, and X. S. Miao, "Work function contrast and energy band modulation between amorphous and crystalline Ge2Sb2Te5 films," *Appl Phys Lett*, vol. 107, no. 8, Aug. 2015, doi: 10.1063/1.4929369.

[51] C. Liu, Y. Wang, J. Liu, R. Ma, H. Liu, Q. Wang, Y. Fu, Q. Liu, and D. He, "Carrier transport mechanisms of titanium nitride and titanium oxynitride electron-selective contact in silicon heterojunction solar cells," *Communications Physics 2024 7:1*, vol. 7, no. 1, pp. 1–9, Jul. 2024, doi: 10.1038/s42005-024-01721-7.

[52] K. Konstantinou and S. R. Elliott, "Atomistic Modeling of Charge-Trapping Defects in Amorphous Ge-Sb-Te Phase-Change Memory Materials," *physica status solidi (RRL) – Rapid Research Letters*, vol. 17, no. 8, Aug. 2023, doi: 10.1002/pssr.202200496.

[53] K. Konstantinou, S. R. Elliott, and J. Akola, "Inherent electron and hole trapping in amorphous phase-change memory materials: Ge $_2$ Sb $_2$ Te $_5$," *J Mater Chem C Mater*, vol. 10, no. 17, pp. 6744–6753, 2022, doi: 10.1039/D2TC00486K.

[54] K. Konstantinou, F. C. Mocanu, J. Akola, and S. R. Elliott, "Electric-field-induced annihilation of localized gap defect states in amorphous phase-change memory materials," *Acta Mater*, vol. 223, p. 117465, Jan. 2022, doi: 10.1016/j.actamat.2021.117465.

[55] X. Yang, W. Liu, M. De Bastiani, T. Allen, J. Kang, H. Xu, E. Aydin, L. Xu, Q. Bi, H. Dang, E. AlHabshi, K. Kotsovos, A. AlSaggaf, I. Gereige, Y. Wan, J. Peng, C. Samundsett, A. Cuevas, and S. De Wolf, "Dual-Function Electron-Conductive, Hole-Blocking Titanium Nitride Contacts for Efficient Silicon Solar

Cells," *Joule*, vol. 3, no. 5, pp. 1314–1327, May 2019, doi: 10.1016/j.joule.2019.03.008.

[56] D. Ielmini, "Threshold switching mechanism by high-field energy gain in the hopping transport of chalcogenide glasses," *Phys Rev B Condens Matter Mater Phys*, vol. 78, no. 3, Jul. 2008, doi: 10.1103/PhysRevB.78.035308.

[57] M. T. Bin Kashem, J. Scoggin, H. Silva, and A. Gokirmak, "(Digital Presentation) Finite Element Modeling of Thermoelectric Effects in Phase Change Memory Cells," *ECS Trans*, vol. 108, no. 1, p. 3, May 2022, doi: 10.1149/10801.0003ECST.

[58] L. Adnane, N. Williams, H. Silva, and A. Gokirmak, "High temperature setup for measurements of Seebeck coefficient and electrical resistivity of thin films using inductive heating," *Review of Scientific Instruments*, vol. 86, no. 10, Oct. 2015, doi: 10.1063/1.4934577/15874753/105119_1_ACCEPTED_MANUSCRIPT.PDF.

[59] M. Tashfiq, B. Kashem, R. Sayeed Khan, A. Hasan Talukder, F. Dirisaglik, and A. Gokirmak, "Stopping Resistance Drift in Phase Change Memory Cells and Analysis of Charge Transport in Stable Amorphous Ge2Sb2Te5," *arXiv: 2210.14035*, Oct. 2022, Accessed: Jan. 26, 2025. [Online]. Available: https://arxiv.org/abs/2210.14035v1.

[60] J. L. Lin and S. C. Lee, "Amorphous silicon thin film transistors," *Journal of the Chinese Institute of Engineers*, vol. 18, no. 4, pp. 451–460, 1995, doi: 10.1080/02533839.1995.9677710.

[61] J. Becker, E. Fretwurst, and R. Klanner, "Measurements of charge carrier mobilities and drift velocity saturation in bulk silicon of <111> and <100> crystal orientation at high electric fields," *Solid State Electron*, vol. 56, no. 1, pp. 104–110, Jul. 2010, doi: 10.1016/j.sse.2010.10.00.

[62] P. M. Smith, M. Inoue, and J. Frey, "Electron velocity in Si and GaAs at very high electric fields," *Appl Phys Lett*, vol. 37, no. 9, pp. 797–798, 1980, doi: 10.1063/1.92078.

[63] A. Cywar, J. Li, C. Lam, and H. Silva, "The impact of heater-recess and load matching in phase change memory mushroom cells," *Nanotechnology*, vol. 23, no. 22, Jun. 2012, doi: 10.1088/0957-4484/23/22/225201.

[64] R. R. Shanks, "Ovonic threshold switching characteristics," *J Non Cryst Solids*, vol. 2, pp. 504–514, Jan. 1970, doi: 10.1016/0022-3093(70)90164-X.

[65] J. Scoggin, A. Gokirmak, R. S. Khan, Z. Woods, A. Cywar, L. Adnane, S. Muneer, G. Bakan, F. Dirisaglik, and H. Silva, "Electrothermal Modeling of Ovonic Threshold Switches and Phase Change Memory Devices," 2019.

[66] W. Czubatyj and S. J. Hudgens, "Invited paper: Thin-film Ovonic threshold switch: Its operation and application in modern integrated circuits," *Electronic Materials Letters*, vol. 8, no. 2, pp. 157–167, Apr. 2012, doi: 10.1007/s13391-012-2040-z.

[67] J. Scoggin, H. Silva, and A. Gokirmak, "Field dependent conductivity and threshold switching in amorphous chalcogenides—Modeling and simulations of ovonic threshold switches and phase change memory devices," *J Appl Phys*, vol. 128, no. 23, Dec. 2020, doi: 10.1063/5.0027671.

[68] M. P. West, G. Pavlidis, R. H. Montgomery, F. F. Athena, M. S. Jamil, A. Centrone, S. Graham, and E. M. Vogel, "Thermal environment impact on HfOx RRAM operation: A nanoscale thermometry and modeling study," *J Appl Phys*, vol. 133, no. 18, May 2023, doi: 10.1063/5.0145201.

[69] J. J. Yang, M. D. Pickett, X. Li, D. A. A. Ohlberg, D. R. Stewart, and R. S. Williams, "Memristive switching mechanism for metal/oxide/metal nanodevices," *Nat Nanotechnol*, vol. 3, no. 7, pp. 429–433, 2008.

[70] J. Dallas, G. Pavlidis, B. Chatterjee, J. S. Lundh, M. Ji, J. Kim, T. Kao, T. Detchprohm, R. D. Dupuis, S. Shen, S. Graham, and S. Choi, "Thermal characterization of gallium nitride p-i-n diodes," *Appl Phys Lett*, vol. 112, no. 7, p. 073503, Feb. 2018, doi: 10.1063/1.5006796.

[71] A. Deschenes, S. Muneer, M. Akbulut, A. Gokirmak, and H. Silva, "Analysis of self-heating of thermally assisted spin-transfer torque magnetic random access memory," *Beilstein Journal of Nanotechnology*, vol. 7, no. 1, 2016, doi: 10.3762/bjnano.7.160.

[72] B. D. Josephson, "Possible new effects in superconductive tunnelling," *Physics Letters*, vol. 1, no. 7, pp. 251–253, Jul. 1962, doi: 10.1016/0031-9163(62)91369-0.